\documentstyle[12pt, aaspp4]{article}
\received{}
\revised{}
\accepted{}
\slugcomment{}
\lefthead{GORHAM et al.}
\righthead{MKN 421 COMPANION}
\begin{document}
\title{Markarian 421's Unusual Satellite Galaxy}
\vspace{0.5cm}
\author{ Peter~W.~Gorham \altaffilmark{1}, 
Liese~van Zee  \altaffilmark{2,}\altaffilmark{3} 
Stephen~C.~Unwin \altaffilmark{1}, and
 Christopher~S.~Jacobs \altaffilmark{1}}

\vspace{0.3cm}
\affil{1. Jet Propulsion Laboratory, \\
Pasadena, California, 91109}
\vspace{0.3cm}
\affil{2. National Radio Astronomical Observatory, \\
PO Box 0, Socorro, New Mexico}
\altaffiltext{3}{presently at: 
Dominion Astrophysical Observatory, 
5071 West Saanich Road, \\
Victoria, B.C., V8X 4M6, Canada}
\vspace{0.3cm}

\begin{abstract}
We present Hubble Space Telescope (HST) imagery and photometry
of the active galaxy Markarian 421 and its  
companion galaxy 14 arcsec to the ENE. The HST images indicate
that the companion is a morphological spiral rather than
elliptical as previous ground--based imaging has concluded. The 
companion has a bright, compact nucleus, 
appearing unresolved in the HST images.
This is suggestive of Seyfert activity, or 
possibly a highly luminous compact star cluster.

We also report the results of high dynamic range long-slit spectroscopy
with the slit placed to extend across both galaxies and nuclei.
We detect no emission lines in the companion nucleus, though there
is evidence for recent star formation. Velocities derived
from a number of absorption lines visible in both galaxies
indicate that the two systems are probably tidally bound and thus in
close physical proximity. Using the measured relative velocities,
we derive a lower limit on the MKN 421 mass within the companion orbit
($R \sim 10$ kpc) of $5.9 \times 10^{11}$ solar masses, 
and a mass--to--light ratio of $\ge 17$. 

Our spectroscopy also shows for the first time the presence of
H$\alpha$ and [NII] emission lines from the nucleus of MKN 421,
providing another example of the appearance of
new emission features in the previously featureless spectrum
of a classical BL Lac object. We see both broad and narrow line emission,
with a velocity dispersion of several thousand km s$^{-1}$ 
evident in the broad lines.

\end{abstract}

\keywords{galaxies: BL lacertae objects: individual (Markarian 421) ---
galaxies: binary --- galaxies: interactions }

\section{Introduction}

Markarian 421 is a giant elliptical galaxy that contains one of the
nearest BL Lac objects, at a redshift of 0.0308 (Ulrich et al. 1975). 
This object is among the most intensively
studied of all active galactic nuclei (AGN). 
MKN 421 is a strong cm-wavelength radio source
that has shown reported superluminal motion in its compact radio jet (Zhang 
and Baath 1990), although recent measurements (Piner et al. 1999) do not
confirm this.
It is optically
highly variable in both intensity and polarization.
It is seen in X-rays (Comastri et al. 1997), 
GeV  gamma-rays (Mukherjee et al. 1997), 
and up to multi-TeV energies (Krennrich et al. 1997). Episodes of rapid
variability have been seen repeatedly at many wavelengths
(Tosti et al. 1998; Gaidos et al. 1996)
strengthening the evidence for
the presence of a compact object as the source of the nuclear activity.

The host galaxy of MKN 421 has been the subject of several spectroscopic 
and photometric studies. The first such study (Ulrich 1975) established
the redshift based on weak stellar absorption lines,
and also noted that a nearby galaxy 14 arcsec to the ENE had a similar
redshift (z=0.0316), indicating that it was probably physically related,
although if the velocity difference were due to the Hubble flow, the
distance could be a few Mpc or more. The companion galaxy was classified as
a normal elliptical (Hickson et al. 1984).
Further work by Ulrich (1978) showed that MKN 421 was the brightest member
of a group of 5--7 galaxies of similar redshift spread over a region of
sky of order 10 arcmin in radius. The presence of this group
increases the likelihood that the companion's proximity to MKN 421 is
physical rather than a random alignment.

There is mounting evidence that AGN phenomena appear to be associated with
galaxy mergers or encounters (cf. Shlosman, Begelman, and Frank 1990;
Hernquist \& Mihos 1995). In the case of BL Lac
objects, a significant number have been found in
the last decade or so to be associated with close companions or
groups of nearby galaxies (cf. Falomo 1996; Heidt 1999), although MKN 421
has apparently been overlooked in this regard. 
Intrigued by the proximity of these
two galaxies, we have analyzed archival HST imagery 
of the system, and performed Hale 5 m
long-slit spectroscopic observations aimed at clarifying this
association.
 
Our goal was to understand the nature of the nearby galaxy in relation
to MKN 421, and to investigate the properties of the companion galaxy itself.
If the galaxy is as close to MKN 421 as its projected distance ($\sim 10$ Kpc)
suggests, it is deep within the gravitational potential well of
MKN 421, and is probably sweeping through its stellar halo. The conditions
under which such an encounter can take place are of general interest
in the understanding of galaxy evolution. Our results
will show that the companion galaxy contains a  
Seyfert-like nucleus, and is likely to be tidally interacting with MKN 421. Although the
evidence is circumstantial, this association does appear to lend weight
to the suggestions that galaxy encounters are an important
factor in AGN evolution, and that close companions are associated
in some way with the BL Lac phenomenon.

In the following section we summarize the 
the imagery and photometry which indicate nuclear activity in
the companion galaxy. 
Section 3 summarizes the
spectroscopic results, from which we derive a velocity profile which
indicates a system that is tidally bound.
Section 4 presents some further analysis of the spectroscopic results,
including a derived lower limit of the bulge mass and mass--to--light
ratio of MKN 421 under some plausible assumptions. 
Section 5 summarizes and concludes the article.

\section{Observations}

A complete log of all observations presented here is shown in 
Table \ref{obslog}.

{\small
\begin{table}[h]
\tablenum{1}
\centering
\caption{Log of observations presented here.}
\label{obslog}
\medskip
\centering
\begin{tabular}{lcccc}
\tableline
Date & Telescope & Instrument & Filter/grating  \\ \tableline \\
1997 March 5  & HST & WF/PC2 & F702W \\
1999 February 19 & Palomar 5m & Double Spect. & 600/300 l/mm \\
1999 May 24  & HST & WF/PC2 & F555W \\
1999 May 24  & HST & WF/PC2 & F814W \\
\tableline \\
\end{tabular}
\end{table}
}

 In the work of Ulrich (1978), the system $14\arcsec$ to the ENE 
of MKN 421 was denoted as
galaxy number 5 of the group of galaxies associated with MKN 421. Although
the galaxy had been noticed prior to this work, it appears that
Ulrich was the first to provide a designation for it. 
Thus we will refer to it in this present work
as MKN 421-5. 

We note that users of NASA's Extragalactic Database (NED)
may find a reference to this object as RX J1104.4+3812:[BEV98] 014.
However, since this reference is only used to distinguish
objects within the ROSAT error circle for MKN 421, and the object is 
previously known, we prefer the designation given by Ulrich (1978).

\subsection{HST Imagery and Photometry}

\subsubsection{HST images}

MKN 421 was observed with the HST wide field/planetary camera (WF/PC2)
on 1997 March 5, using the F702W filter. Five exposures, 
one of duration 2 s, two of duration 30 s, and two of duration 120 s were made
\footnote{The HST observations used here are available
as part of the Space Telescope Science Institute public archive, 
and were made originally as a result of a proposal by C. M. Urry.}.
The images were prepared by standard techniques described in Holtzman et al. 
(1995), and the moderate cosmic ray contamination of the frames was 
repaired by hand using linear interpolation.
In the 2 s exposure, the
MKN 421 nucleus is not saturated, but the companion galaxy is underexposed.
The remaining two images overexpose the AGN, but for both MKN 421's
host galaxy and the companion galaxy the exposures provide good signal-
to-noise ratios over the sky background. The pixel scale in these
images is 0.0455 arcsec per pixel, and the resolution of HST at
the mean filter wavelength (690 nm) is $\sim 0.080$ arcsec; thus the
images are sampled just slightly under the Nyquist frequency.

Secondary corrections, including those associated with 
the known gradient in charge transfer efficiency and the 
pixel area variation across the frame (Holtzman  et al. 1995), are not
corrected for in the displayed images, but we have applied
corrections for these effects in the relative photometry presented in
the next section.

MKN 421 was again observed with WF/PC2 on 1999 May 24 with both
the F555W and F814W filters in snapshot mode, with 300s exposures
in each case.\footnote{These HST observations are also available
as part of the Space Telescope Science Institute public archive, 
and were made originally as a result of a proposal by R. Fanti.}
Since these were single frame integrations, the cosmic ray
removal for presentation purposes is more difficult, and the images
do not add significantly to the structure seen in the F702W image.
Thus we do not present these as part of the imagery
here, although we do include results of the photometry performed 
on these additional images in the following section.

\begin{figure}
\plotone{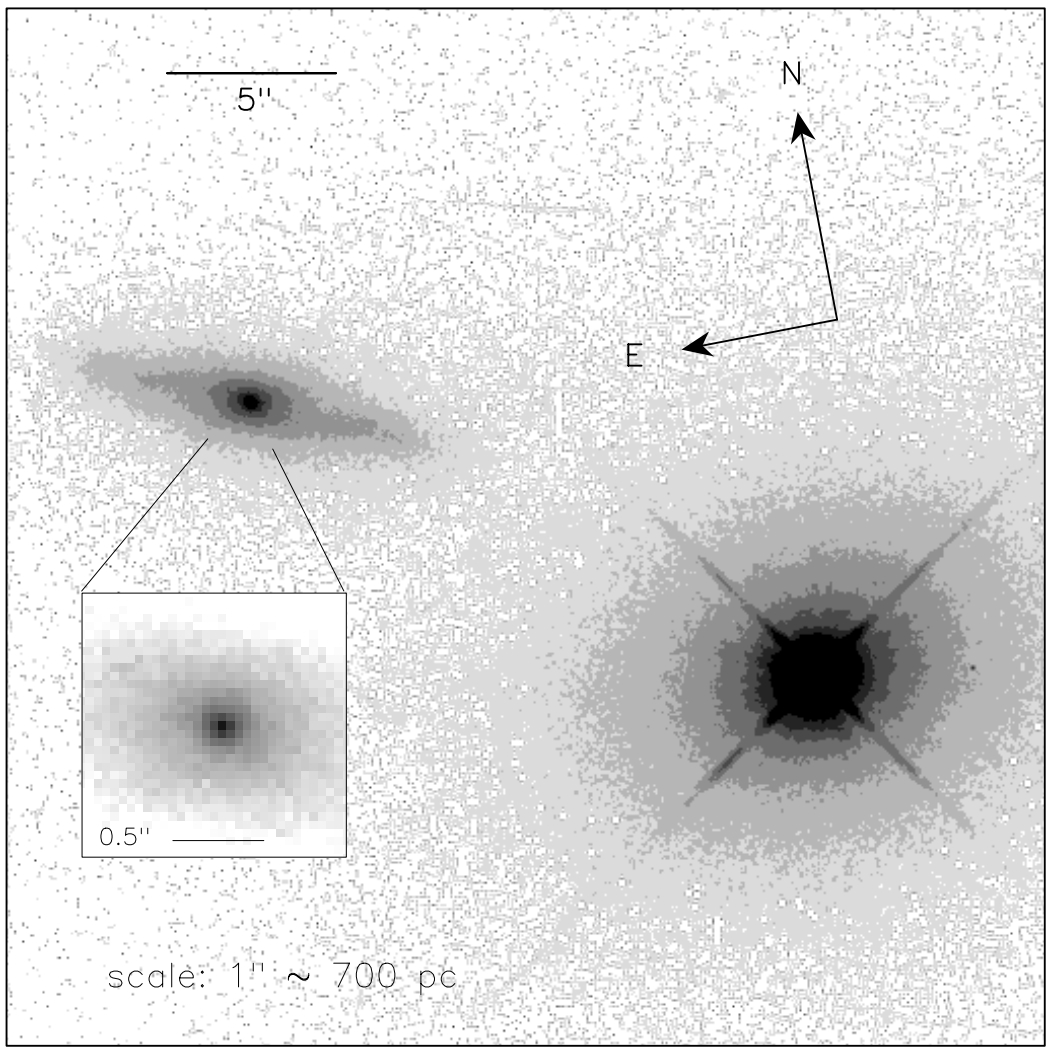}
\figcaption[Gorham.fig1.eps]
{HST image of MKN 421 and companion MKN 421-5, through 
red F702W filter, 300 s total exposure. In the larger image the 
cores of MKN 421 and the companion are saturated to
show detail of the companion disk structure. Levels are quantized
to more clearly show the companion structure.
The inset shows the nucleus of the companion plotted with a 
logarithmic stretch, showing the bright, unresolved nucleus. The
scale is approximately 700 pc/arcsec for $h=0.65$. \label{hst-fig} }
\end{figure}

Figure \ref{hst-fig} shows a slightly smoothed grayscale 
of the summed F702W image, with a logarithmic stretch
and quantized levels chosen 
to show the details of the host galaxy of MKN 421 and MKN 421-5.
Several features of MKN 421-5 are evident even from this image.
First, its structure is not a simple elliptical. A
suggestion of spiral arms is evident, and possible evidence of 
barlike structure appears
at the outer edges of the galaxy. Second, the nucleus of the
companion is clearly brightened relative to the galactic bulge, as is
shown in the inset frame. Third, there is no evidence for any obvious
dust lanes or similar absorption features in either the outer or bulge
regions of the galaxy.

\begin{figure}
\plotone{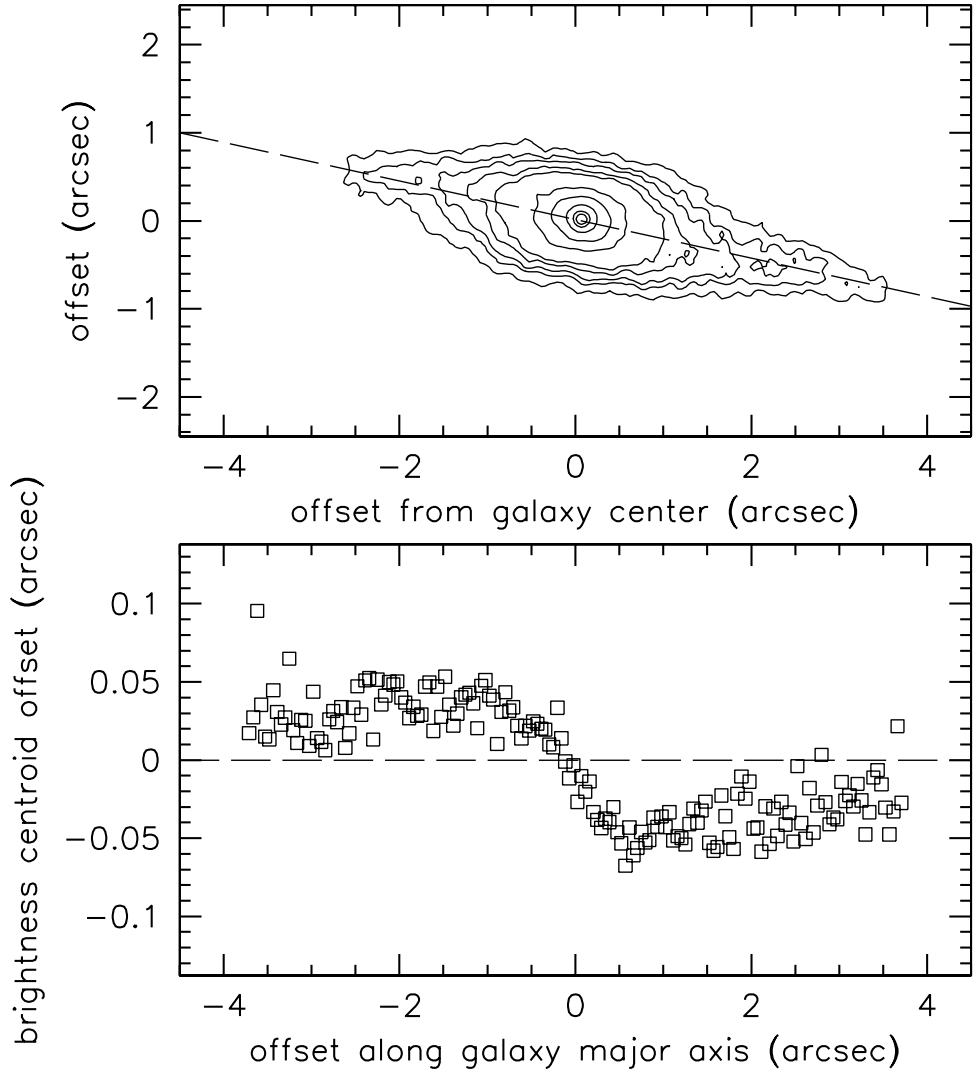}
\figcaption[Gorham.fig2.eps]
{ (a) Logarithmic grayscale of a smoothed HST image 
of MKN 421-5. Central portions 
saturated to show disk more clearly. (b) Plot of surface brightness 
centroids of slices perpendicular to the major axis showing evidence
of spiral structure. \label{companion-fig} }
\end{figure}

\begin{figure}
\plotone{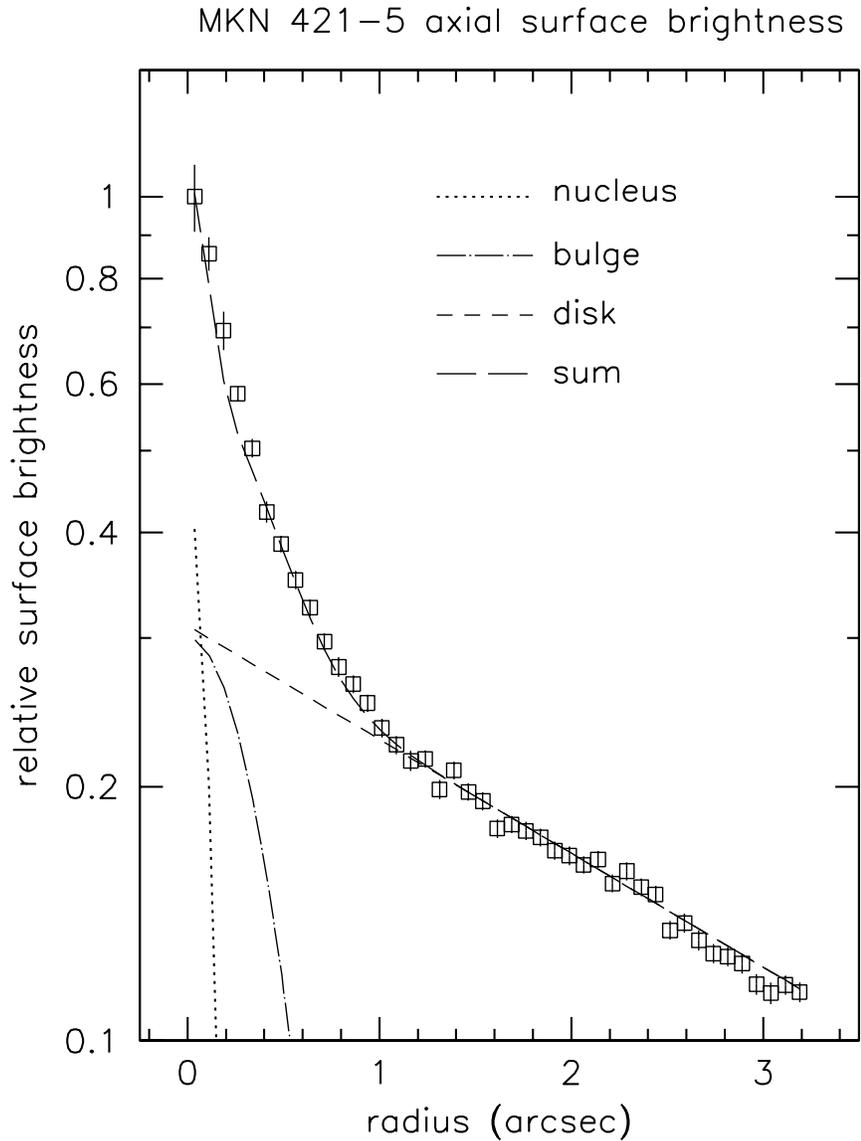}
\figcaption[Gorham.fig3.eps]
{Relative surface photometry of $15^{\circ}$ sectors centered on
the MKN 421-5 major axis, as a function of radial distance from the
galaxy center. We also show best-fit curves for the relative contributions
of an unresolved nuclear source, a surrounding bulge (taken to be a 
Gaussian), and an exponential disk (with no inner truncation). The
need for mutiple components in the fit strengthens the case that this
is actually a disk system. \label{surfphot-fig}}
\end{figure}

Figures \ref{companion-fig}--\ref{surfphot-fig} illustrate these
features. In Figure \ref{companion-fig}  
we show more quantitative evidence for the
presence of spiral structure, since the structure is difficult to reproduce
adequately in paper copies of the images themselves. 
Fig. \ref{companion-fig}(a)
is a contour plot centered on the companion galaxy showing
structure which is suggestive of spiral arms. 
In Fig. \ref{companion-fig}(b) we show centroids of
the surface brightness distributions in slices along the galaxy
major axis (dashed line)
in Fig. \ref{companion-fig}(a). The displacement
of the brightness centroids clearly shows the presence of spiral structure.
(The amplitude of the centroid displacement does not directly
track the location of the arms, since it depends on the brightness
of the arm structure relative to the surrounding disk.)

In Figure \ref{surfphot-fig}
we show a profile of relative surface photometry using the F702W
filter along the major
axis of the companion galaxy. We have averaged the surface brightness in
opposite $\simeq 15^{\circ}$ sectors, and have performed one-dimensional
fits to 3 components that are apparent in the image: an unresolved
core; a bulge component, here represented by a Gaussian; and
an exponential disk. We have not truncated the disk at an inner radius here;
our goal is primarily to show that the data are not consistent with
a single power-law or other one-parameter model for radial brightness,
but are reasonably well-modeled by the standard parameters of 
a disk system.

\subsubsection{HST photometry}

We estimated photometric parameters from the HST images using the
prescription given by Holtzman et al. 1995, including all of the
relevant corrections. We have also transformed the resulting magnitudes
to Cousins {\it VRI} magnitudes using the transformations in Holtzman et al.,
and we include also the transformed colors although we note that
the $V$ and $I$ measurements were not contemporaneous 
with the $R$ measurements. 
In Table \ref{hstphot} we summarize the results for the nuclear magnitudes
of the two galaxies, and the annular bulge magnitude of MKN 421-5.
We have also converted the
measured magnitudes to absolute magnitudes using $h=0.65$, for use
in later discussion.
For MKN 421, we show only an estimate of the $R$ magnitude of the
nucleus for comparison with MKN 421-5; in the 1999 May observations
in the F555W and F814W filters
there were no exposures short enough to avoid saturation of the nucleus.

{\small
\begin{table}[h]
\tablenum{2}
\centering
\caption{HST photometry of MKN 421 \& companion. Units are magnitudes,
and standard errors are typically 0.04 mag. No unsaturated $V$ or $I$ band
exposures of the nucleus of MKN 421 were available.}
\vspace{0.3cm}
\label{hstphot}
\medskip
\centering
\begin{tabular}{lcccccc}
\tableline
Source & $V$  & $R$  & $I$ & $V-R$ & $V-I$ & M$_{R}$  \\ \tableline \\
MKN 421 AGN& ... & 12.74  & ... & ... & ... & -23.0 \\
MKN 421-5 nucleus ($r\leq 0.18$'')& 19.66 & 19.11 & 18.53 & 0.56 & 1.14 & -16.6 
\\
MKN 421-5 bulge ($0.18''\leq r \leq 1.1$'')& 19.62 & 19.05 & 18.41 & 0.58& 1.21 
& -16.7\\
\tableline \\
\end{tabular}
\end{table}
}

Based on the expected contributions of the
different components indicated by Fig. \ref{surfphot-fig},
The nuclear magnitude may contain a significant bulge contribution,
of order 30-40\%.
In spite of this, it still appears that the bright nucleus is
too compact to be a typical star forming region. As we discuss
in a later section, it is likely to be either a compact nuclear star 
cluster or more probably an active nucleus.
And although there is only a marginal difference in color between the
nucleus and the annular bulge region, it does favor a rise of the
nucleus relative to the bulge 
in the near--infrared, which is consistent with Seyfert--nucleus
behavior. 

The nucleus of MKN 421 itself
is highly variable in the visual bands. In fact, the
HST observations in early 1997 occurred shortly after a large optical outburst
(Tosti et al. 1998) during which the R-band magnitude peaked at brighter
than 12. During March 1997, Tosti et al. estimated $R=12.4$ from
ground-based photometry. The HST observations, taken approximately
2 months later, show a decrease of $0.34$ mag.

\section{Optical Spectroscopy}

Low resolution optical spectra of MKN 421 and its companion
were obtained with the Double Spectrograph on the 5m
Palomar\footnote{These observations at the Palomar Observatory 
were made as part of a continuing cooperative agreement
between Cornell University and the California Institute 
of Technology.} telescope during the night of 1999 February 19.
The long slit (2\arcmin) with a 2\arcsec~aperture
was centered on the companion galaxy and two 600 sec
exposures were obtained.  The slit was positioned at
an angle of 53\arcdeg, and passed through both the
companion galaxy and the nucleus of MKN 421.
A 5500 \AA~dichroic was used to split the light to the two 
sides (blue and red), providing nearly complete spectral 
coverage from 3600--7600 \AA.  The blue spectra were acquired
with the 600 lines/mm diffraction grating (blazed at 4000 \AA).
The red spectra were acquired with the 316 lines/mm diffraction
grating (blazed at 7500 \AA).  Thinned 1024$\times$1024 Tek CCDs,
with read noises of 8.6 e$^-$ (blue) and 7.5 e$^-$ (red), were
used on the two sides of the spectrograph.  Both CCDs had
a gain of 2.\ e$^-$/(digital unit).  The effective spectral resolution of
the blue camera was 5.0 \AA~(1.72 \AA/pix); the effective spectral
resolution of the red camera was 7.9 \AA~(2.47 \AA/pix).
The spatial scale of the long slit was 0.62 \arcsec/pix for the
blue camera and 0.48 \arcsec/pix for the red camera.

{\small
\begin{figure}
\plotone{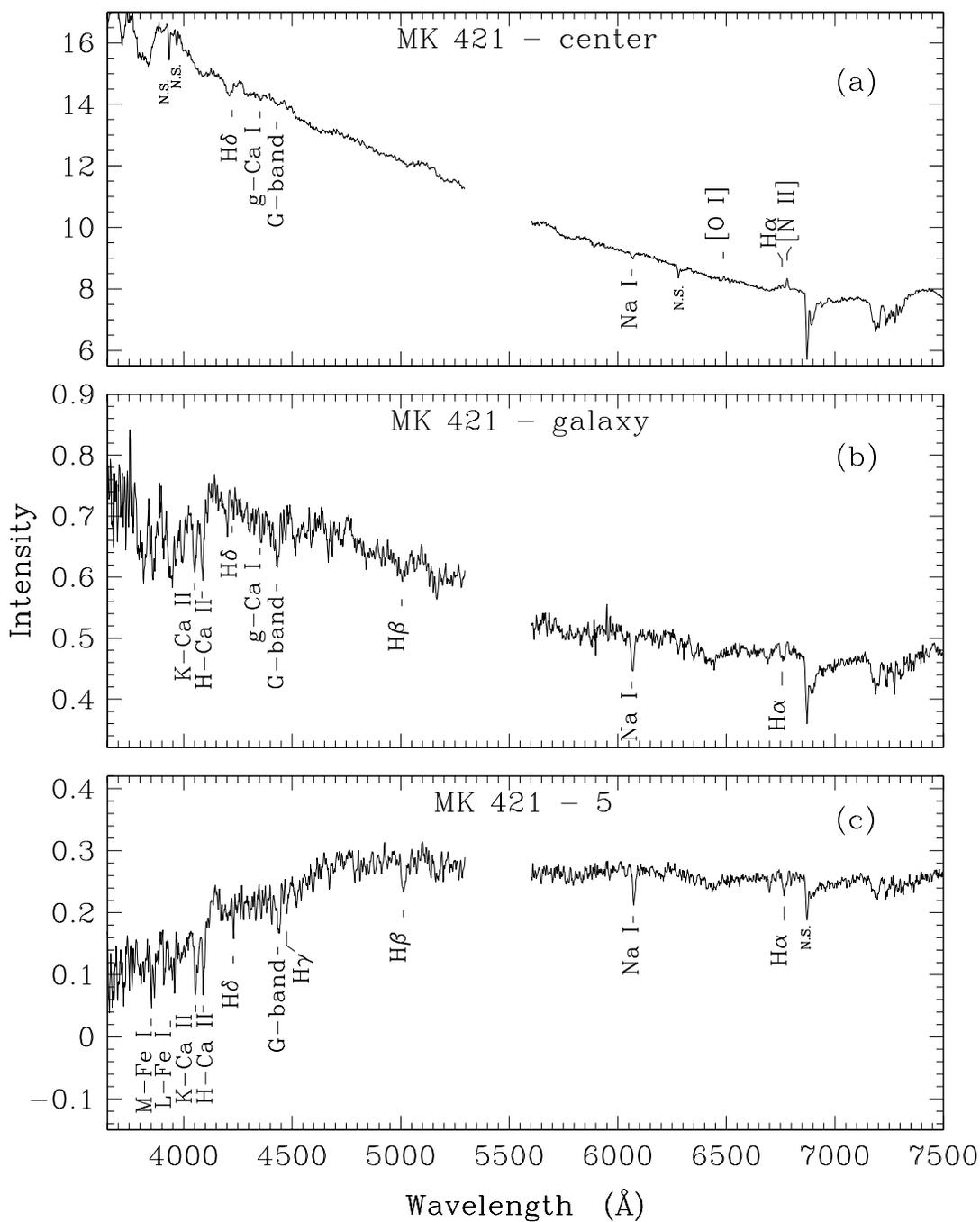}
\figcaption[Gorham.fig4.eps]
{Palomar 5 m double spectrograph spectrum of MKN 421 and
its companion. The slit was aligned with the two nuclei.
Units are  erg cm$^{-2}$ s$^{-1}$ \AA$^{-1}$ with an overall uncertainty
of 20\% in the flux calibration.
(a) MKN 421 shows weak emission at
H$\alpha$ and NII, the first time any emission lines have been noted for
this usually almost featureless BL Lac object. 
(b),(c) MKN 421 and its companion
galaxy have a number of well-defined absorption lines but there is no
detectable emission from either galaxy. \label{spectrum-fig} }
\end{figure}
}
	The spectra were reduced and analyzed with the IRAF\footnote{IRAF 
is distributed by the National Optical Astronomy Observatories.} package.
The spectral reduction included bias subtraction, scattered light
corrections, and flat fielding with both twilight and dome flats.
The 2--dimensional images were rectified based on arc lamp
observations (Fe and Ar for the blue side; He, Ne, and Ar for
the red side) and the trace of stars at different positions along 
the slit.   The sky background was removed from the 2-dimensional
images by fitting a low order polynomial along each row of the
spectra.  One dimensional spectra of MKN 421 and its companion
were extracted from the rectified images using a 1.5\arcsec~extraction 
region (the seeing disk at the time of the observations)
centered on the peak emission of each system.
In addition, a galaxy spectrum for MKN 421 was obtained by averaging
together two 4\arcsec~regions offset from the nucleus by $\pm$4\arcsec.
While the night was non--photometric, observations of
standard stars from the list of Oke \markcite{Oke90}(1990) provided 
calibration for the fluxes which are estimated to be
accurate to $\sim 10$\%. The flux--calibrated spectra are presented 
in Figure \ref{spectrum-fig}, in units of erg cm$^{-2}$ s$^{-1}$
\AA$^{-1}$. We estimate an overall uncertainty of 20\% in the flux
calibration due to variations in the transparency during our observations.

\subsection{Spectrum of MKN 421}

\begin{figure}
\plotone{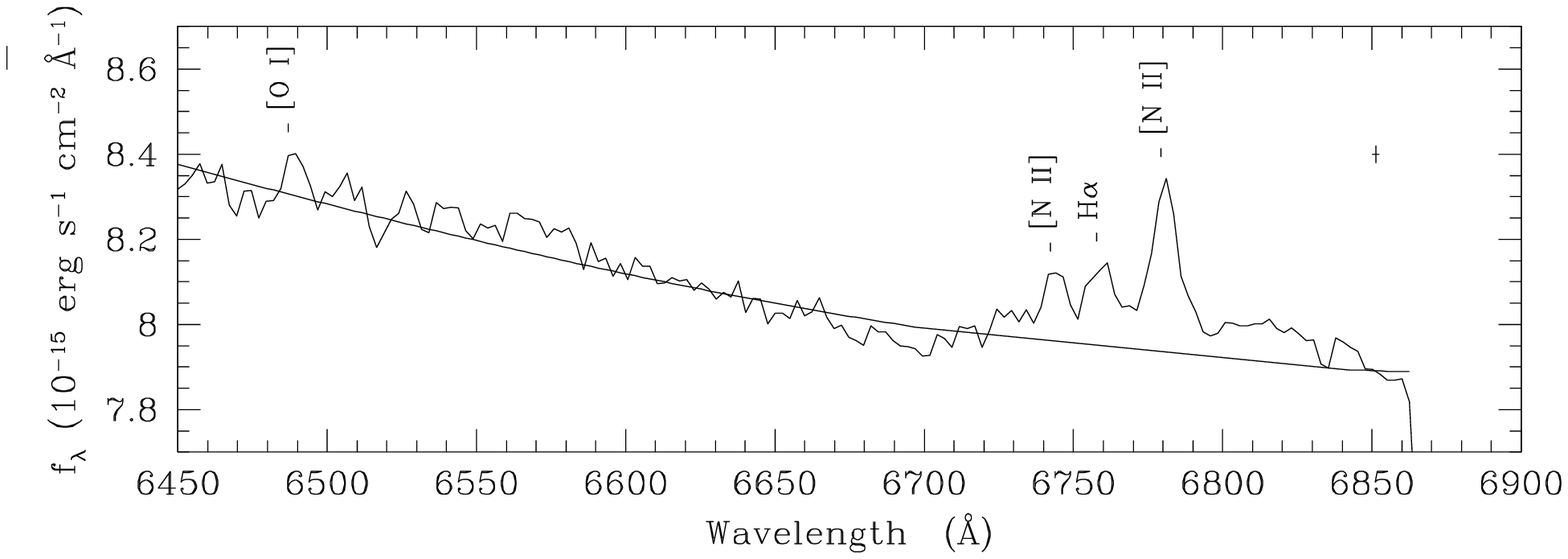}
\figcaption[Gorham.fig5.eps]
{Emission lines seen in MKN 421. The curve shows
a power-law continuum model fitted to the overall spectrum. 
Both broad and narrow emission line
components are evident above the continuum. 
A typical statistical error bar for the individual points is
shown at the upper right. \label{broad-fig}}
\end{figure}

As seen in Figure \ref{spectrum-fig}a, the nucleus of MKN 421 is dominated
by nonthermal emission with very weak absorption features.
This new spectrum is similar to other observations of the nucleus
of MKN 421 (e.g., March\~a et al.\ \markcite{Me96}1996), with the
exception that [OI], [NII], and H$\alpha$ emission lines have been
detected.  While measurement of relative 
fluxes for the narrow emission features is complicated by the 
presence of broad H$\alpha$ emission (Figure \ref{broad-fig}), and by
the contamination of H$\alpha$ absorption from the underlying
stellar population (see Figure \ref{spectrum-fig}b), the [NII] lines are 
significantly stronger than the narrow H$\alpha$ feature.
Large [NII]/H$\alpha$ ratios are not uncommon in AGN, however 
(e.g., Veilleux \& Osterbrock \markcite{VO87}1987).  
Both the broad and narrow emission lines appear
to be associated only with the nucleus of MKN 421.

In Fig. \ref{broad-fig} we plot a fitted continuum model for
MKN 421 along with the measured spectrum, in the region near the
detected emission lines. Although the 
continuum model does not account for all of the structure in
the spectrum near the emission lines, it does qualitatively indicate the
instrinsic width of the broad-line emission, which has
a full-width-at-half-maximum (FWHM) of order 80 \AA~
with detectable emission that 
extends out to nearly twice this value.
The implied velocity dispersion is $\sim 5000$ km s$^{-1}$.
We discuss some of the implications of this in a later section.

Recently, Morganti et al.\ \markcite{Me92}(1992) obtained 
narrow band images roughly centered on the H$\alpha$ and [NII]
lines and reported a total H$\alpha +$[NII] flux of 1.6 $\times$ 
10$^{-14}$ erg s$^{-1}$ cm$^{-2}$ for MKN 421.  
The present observations are
the first spectroscopic detection of emission lines associated 
with the nucleus of MKN 421, confirming this result.  The total 
integrated flux in the broad and narrow emission lines is
1.7 $\times$ 10$^{-14}$ erg s$^{-1}$ cm$^{-2}$ (EW $\sim$ 2.2 \AA).
We estimate an overall uncertainty in this value of $\sim 20$\%.

This apparent agreement with Morganti et al. is somewhat
misleading, since our estimate includes flux spread over 
more than 100 \AA, while the Morganti et al. estimate was based
on a 50 \AA ~FWHM filter which was apparently offset with respect
to the centroid of the emission. Thus we actually have detected
a significantly {\em lower} total flux level than Morganti et al.,
although we cannot quantify the difference without a more accurate
knowledge of the filters used by Morganti et al. 
Despite the new detection of emission lines, MKN 421
still falls under the general class of BL Lac objects:
the 4000 \AA~break has a low contrast in the nuclear
regions and the derived equivalent widths for the emission 
lines are significantly less than 5 \AA. 

The sudden appearance of emission features in previously
featureless spectra of BL Lac--type objects is becoming
quite common.  Just a few years ago, such features were
detected in the prototypical BL Lac, BL Lac itself,
for the first time (Vermeulen et al.\ \markcite{Ve95}1995;
Corbett et al.\ \markcite{Ce96}1996).  Similar events
were discovered in OJ 287 (Sitko \& Junkkarinen 
\markcite{Sitko85}1985) and PKS 0521--365 (Ulrich 
\markcite{U81}1981; Scarpa, Falomo, \& Pian \markcite{Scarpa95}1995) 
as well.  The rise and decline of emission--line features
in BL Lac--type objects poses an interesting puzzle,
and monitoring of these lines should be done as often as
possible to determine
if their intensity is correlated to that of other bands.

\subsection{Spectrum of the Host Galaxy}

A spectrum of the underlying host galaxy was obtained
by averaging spectra on either side of the nucleus
(Figure \ref{spectrum-fig}b). As found in other spectroscopic observations 
of the host galaxy (e.g., Ulrich et al.\ \markcite{Ue75}1975; 
Ulrich \markcite{U78}1978), the spectrum is dominated
by absorption features. In addition to the usual strong
absorption features in the blue, the Na I $\lambda\lambda$5890,5896
doublet is remarkably strong throughout the galaxy.
We note that the presence of an apparent broad emission bump in the 
region of the spectrum surrounding the H$\alpha$ absorption line is
apparently due to scattered emission from the extremely bright
active nucleus of MKN 421.

\subsection{Spectrum of MKN 421-5}

Similar to the host galaxy of MKN 421, the spectrum
of the companion galaxy is dominated by absorption
lines (Figure \ref{spectrum-fig}c).  
However, the presence of strong Balmer absorption
features indicates that this system is not dominated only
by an old stellar population; rather, it must have had recent
star formation activity within the last Gyr or so.

The plethora of absorption lines in the optical spectrum
permit an accurate redshift determination for this system: 
9380 $\pm$ 50 km s$^{-1}$.  Ulrich \markcite{U78}(1978) was
one of the first to point out that MKN 421 is part of a large 
group of galaxies; we now know that radio galaxies tend to 
form in groups (Zirbel \markcite {Z97}1997).  The close
proximity (spatially and in velocity space) of MKN 421 and
MKN 421-5 suggests that these two systems may 
have had significant tidal interactions in the past and
may explain the unusual stellar population of the companion.

\subsection{Spatially resolved Na absorption}

\begin{figure}
\plotone{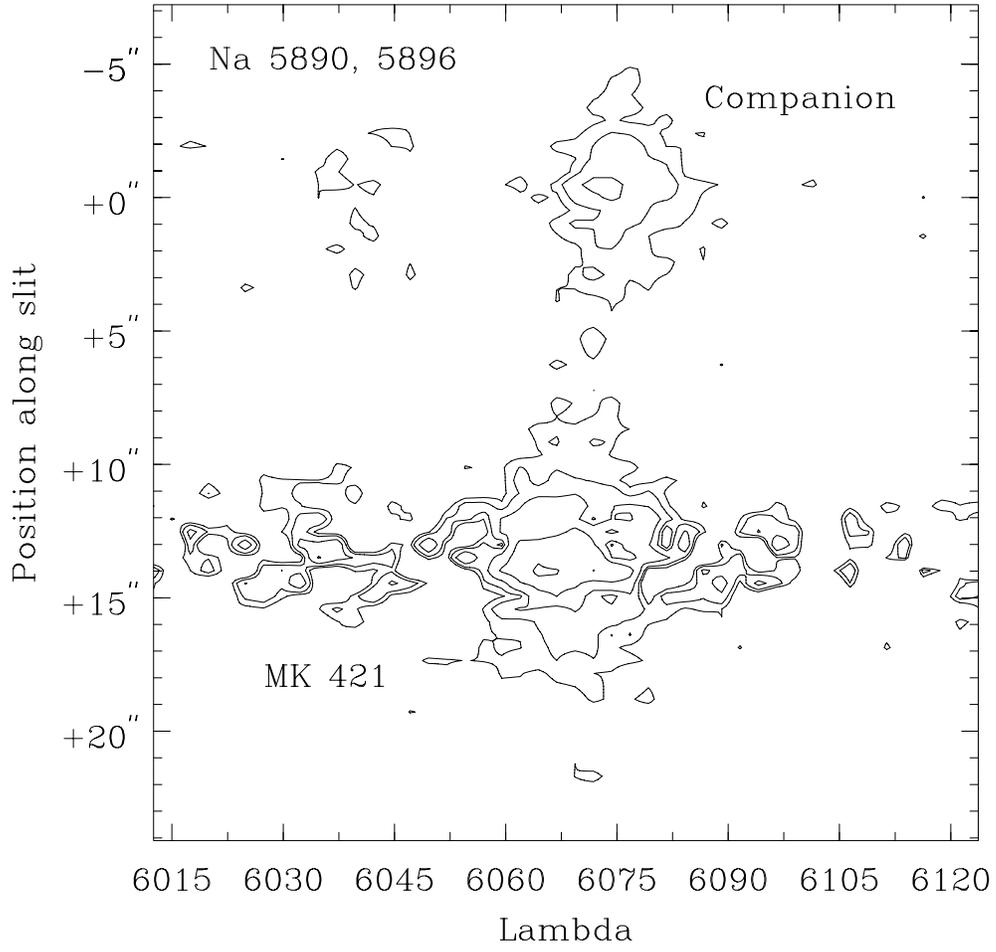}
\figcaption[Gorham.fig6.eps]
{Spatially resolved absorption line profile 
for the NaI doublet. The  centroid of the doublet shows a velocity trend 
that smoothly joins the two galaxies, an indication of a 
tidal interaction. \label{Naprofile-fig}}
\end{figure}

\begin{figure}
\plotone{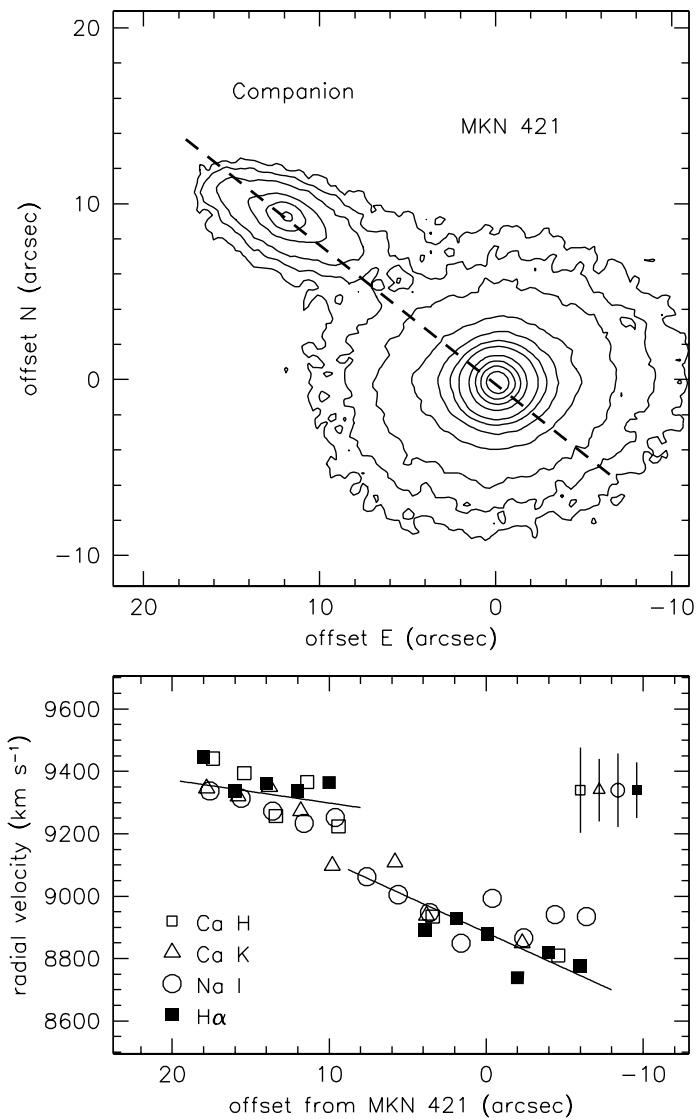}
\figcaption[Gorham.fig7.eps]
{(a) Contour plot of Palomar 60'' Gunn r-band image of
MKN 421 and companion. The heavy dashed line marks the slit
position for the absorption line spectroscopy.
(b) Plot of line centroids as a function of slit position. 
The lines show the
fitted velocity gradients across each galaxy separately. The
velocity gradient of MKN 421 is consistent with a component of Keplerian
rotation that is in the same sense as the companion's velocity relative
to MKN 421. The two thus appear to be a bound pair. \label{gradvel-fig}}
\end{figure}

Both MKN 421 and its companion galaxy have high signal--to--noise
Na I absorption features which can be used to trace their
gas kinematics.  A position--wavelength diagram for the two 
systems is shown in Figure \ref{Naprofile-fig}.  The Na line can be traced
almost continuously from MKN 421 to MKN 421-5.
The mean velocity as a function of position along the slit 
for several of the absorption lines is shown in Figure \ref{gradvel-fig}.
The errors in each individual measurement are largely 
due to the relatively low signal--to--noise ratio 
and the somewhat coarse spectral resolution of the 
observations; nonetheless, the two systems are clearly
offset in velocity, with a sense of velocity continuity
between the two galaxies.

\placefigure{gradvel-fig}

{\small
\begin{table}[h]
\tablenum{3}
\centering
\caption{Fitted velocities and velocity gradients for MKN 421 \& companion.}
\vspace{0.3cm}
\label{vgrad}
\medskip
\centering
\begin{tabular}{lcccc}
\tableline \\
Source & $\bar{v}$ & $\sigma_{v}$ & $\nabla v$ & $\sigma_{\nabla v}$ \\
  &km s$^{-1}$ &km s$^{-1}$ &km s$^{-1}$ as$^{-1}$ & km s$^{-1}$ as$^{-1}$  \\
 \tableline \\
MKN 421 &  8884 & 27 & 23.0 & 6.0 \\
MKN 421-5 & 9380 & 50 & 7.4 & 4.0 \\ \tableline \\
\end{tabular}
\end{table}
}

Fig. \ref{gradvel-fig} also displays the results 
of unconstrained least-squares fits
for the velocity gradients across each of the two systems. The results of
these fits are shown in Table \ref{vgrad}. The velocities
are not corrected for earth orbital motion; however, the correction for
the hour angle of our observation is 
negligible compared to the errors in the estimates in Table \ref{vgrad}.

\subsection{On limits to radio emission from MKN 421-5}

A number of VLA studies of MKN 421 and its immediate vicinity
were conducted soon after it was recognized to be a BL Lac object
(Ulvestad et al. 1984; 1983); however, the resolution of these
studies was not adequate to show the companion due in part to the high
dynamic range required (MKN 421 is $0.57$ Jy at 20cm). More
recent 20cm observations (Laurent-Muhleisen et al. 1993) with higher resolution
show complex extended emission in the vicinity of MKN 421, 
but probably do not have enough dynamic range to detect the
companion galaxy nucleus, if its brightness is comparable to
the average of its class.

Although there are conspicuous examples of radio--loud Seyferts
(primarily Seyfert 1) more typical
Seyfert nuclei of both type 1 and 2 emit total fluxes of order
$10^{21}$ W Hz$^{-1}$ at cm wavelengths (cf. Ulvestad and Wilson 1989 and
references therein). At the distance of the companion,
this luminosity would produce about 0.5 mJy of 3.5 cm flux density,
and a factor of 2-3 more than this at 20 cm for a typical
Seyfert spectral index of $-0.5$ to $-0.7$.
Inspection of a number of VLA maps at various resolutions shows no
source of significant strength at the companion galaxy position.
However, the complex halo of emission around MKN 421 precludes
identifying a source at a level below 1 mJy. Further
radio observations should be made to attempt to identify any
compact source associated with the companion galaxy nucleus.

\section{Discussion}

\subsection{The nature of the companion}

The absolute magnitude of the nucleus of MKN 421-5 was given in Table 2
as $M_R = -16.6$. If we assume this magnitude contains 
a 40\% background contribution from the
galaxy of MKN 421-5, the nucleus still has $M_R = -16.0$, and 
a corresponding $M_V \simeq -15.5$.
It is thus apparently as bright or brighter than any of the compact nuclear
clusters described from HST observations by Phillips et al. (1996)
and Carollo et al. (1997) which fell mostly 
in the range of $M_V = -12$ to $-14$.
A recent detailed HST study of four
nearby spirals with compact nuclei 
(Matthews et al. 1999) gave $M_B$ values that
range from -8.5 to -10.4, and corresponding $M_I$ values of
-9.9 to -11.4. 

Compact star cluster nuclei are not uncommon among late-type spiral galaxies,
although they are often difficult to detect in ground-based images.
However, based on the photometric results presented above, the 
nucleus of MKN 421-5 must be among the most luminous in its class
if it comprises an overluminous nuclear cluster. 

The alternative is that we are viewing a 
Seyfert nucleus near the low end of the Seyfert luminosity range.
In fact, recent studies have shown that the Seyfert luminosity
function extends well below the luminosity of MKN 421-5
(Ho et al. 1996, Ho et al. 1997).
It is also worth noting that Malkan et al. (1998), in an HST
survey targeting known or potential Seyfert galaxies,
found that the presence of a bright, unresolved
nucleus in HST images was a nearly perfect indicator of Seyfert
activity in a survey of several hundred objects. Thus
the lack of emission lines from the nucleus of the companion galaxy is
puzzling, given the other indications that the source is non-thermal in
nature. However, given the mediocre seeing conditions during the
spectroscopic observations, and the extreme compactness of the core, our
constraints on emission lines are not yet very strong. 

The HST results indicate that we are viewing the companion at
a substantial inclination relative to its spiral axis. This could
contribute to the lack of observed emission lines due to
obscuration. If the galaxy is a Seyfert 2, the narrow line emission
may in fact be sufficiently diluted by the galactic
bulge that it will be difficult 
to observe from the ground. From the HST image, the nuclear source is
probably unresolved; even if we assume its diameter to be 0.1'',
the dilution factor under our seeing conditions (1.2'') 
is at least a factor of 10, after accounting for the brightness of
the nucleus relative to its surrounding bulge.
Clearly, interpretation of any ground-based observations of this 
and other similar objects must be tempered with caution because
of the danger of these dilution effects.

\subsection{Implications of the velocity measurements}

\subsubsection{Mass of MKN 421 bulge}

As noted above, the shape of the velocity curve is consistent with
the behavior expected from tidal interaction. This, combined with the
proximity of  MKN 421-5 to MKN 421, and the presence of a group of galaxies
of similar redshift, leads to the conclusion that MKN 421-5 is undergoing
an orbital encounter with MKN 421. Also, since MKN 421-5 is much closer
to MKN 421 than any other galaxy in the group, we may treat the pair
as a binary system for purpose of estimating the dynamical masses involved.

The statistics of binary galaxies have been treated in detail by
Noerdlinger (1975) who showed that the assumption of circular
orbits for a binary pair is a conservative one with regard to mass estimation.
In fact, the circular orbit approximation is valid up to eccentricities
of $\epsilon \sim 0.4$, and underestimates the mass by only a factor 
of 3 in the limit of a parabolic encounter ($\epsilon = 1$).

Thus we  estimate a lower limit to
MKN 421's mass $M$ within the companion's
present projected orbital radius $R_p$:
\begin{equation}
\label{mass-equ}
M(R_p) ~\geq~ {U_r^2 R_p \over G }
\end{equation}
where $U_r$ is the companion's radial velocity and $G$ is the gravitational
constant.

For $U_r = 496$ km s$^{-1}$ and $R_p=10$ Kpc in the comoving frame, the 
lower limit of the mass is
\begin{equation}
M(R_p) ~\geq~ 5.9 \times 10^{11}~ M_{\sun}.
\end{equation}

For moderate orbital eccentricity
($\epsilon \leq 0.4$), Noerdlinger (1975) has
shown that the most probable values of the total velocity
$U_T$ and true separation $R_T$ are likely to be 
20-30\% higher than the 
observed values. Thus the
most probable value for the bulge mass is about twice the value above,
of order $10^{12}~ M_{\sun}$.
Typical giant elliptical rotation curves increase out to radii of
order 10 kpc, then flatten and continue flat out to 50 kpc or more.
Since a flat rotation curve implies a linear increase of the
dark halo mass with radius, we expect that 
only of order 10--20\% of the total mass of the MKN 421 system is likely to 
reside within $R_p$, the total mass of the galaxy probably approaches
$10^{13}~M_{\sun}$, among the highest masses of giant elliptical galaxies. 
Further measurements of the virial velocities of other members of
MKN 421's group should be able to confirm this.

\subsubsection{Mass--to--light ratio}

The CCD photometry of MKN 421 reported by Kikuchi et al.\ (1987) gives the
absolute V magnitude of the MKN 421 host galaxy within a 13\arcsec~radius
of M$_V$(13\arcsec) = --21.51 $\pm$ 0.03, implying a luminosity of 3.5 
$\times$ 10$^{10}$ L$_{\odot}$. Note that the emission from other bands, such
as radio continuum and X--ray, is almost entirely associated with the
AGN rather than the galaxy.  Thus, the implied minimum mass--to--light
ratio is M/L $\geq$ 17, indicating a substantial dark matter contribution
to the total mass of MKN 421 within $R_p$. 

This calculation ignores the possible 
contribution of atomic or molecular gas to the total mass, but no HI has been 
detected in this system (van Gorkom et al. 1989). IRAS measurements
of far-infrared fluxes of MKN 421 (van Gorkom et al.)
have been interpreted as due to the presence of
a total dust mass of order $5 \times 10^{7}$ M$_{\sun}$. If a gas component 
were present corresponding to this dust level, one might expect up to
$10^9$ M$_{\sun}$ of gas in the total galaxy; only a fraction of this
would be within $R_p$. Thus we expect our derived M/L ratio to be
robust. 

A number of statistical estimates of the mass--to--light ratios of binary
galaxies have yielded values with ranges of 12--32 for pure spiral
pairs (Schweizer 1987; Honma 1999) and 22--60 for elliptical pairs 
(Schweizer 1987).  These estimates use galaxies with mean separations
of order 60--100 Kpc, and thus include a much greater mass fraction
of the whole galaxy than our estimate. As discussed above, typical
giant elliptical rotation curves lead one to expect that the total 
M/L for MKN 421 may be 5-10 times higher than what is measured for this
close pair. 

\subsubsection{Estimates of MKN 421 central black hole mass}

Wandel (1999), Wandel et al. (1999), Laor (1998) and others have 
noted the correlation of AGN host galaxy bulge mass $M_{bulge}$
to the mass of a central black hole $M_{BH}$ in cases 
where a reliable virial mass or reverberation
mapping mass could be estimated. Wandel (1999) finds a relation
of $M_{BH} \geq 3 \times 10^{-4}~M_{bulge}$ in a sample 
composed primarily of low-luminosity AGN for which the statistics
are reasonably good. Magorrian et al. (1998), in a comprehensive study of
kinematics of normal galaxies, found a relation 
$M_{BH} \geq 6 \times 10^{-3}~M_{bulge}$. If we use these values to bound the
likely value of $M_{BH}$ for MKN 421, and 
assuming that our estimate of the lower
limit to the mass within $R_p$ is dominated by the bulge mass of MKN 421, 
we expect a central black hole mass in the range $1.8 \times 10^8
\leq M_{BH} \leq 3.6 \times 10^9$ M$_{\sun}$.

We can also use the parametric relations given by Laor (1998) 
and Wandel et al. (1999) to make an independent estimate of the
size of the broad-line region (BLR) based on the AGN bolometric luminosity.
Wandel et al. give a black hole mass
estimate based on the size of the BLR and $\Delta v_{FWHM}$, the FWHM of the
measured broad-line velocity dispersion:
\begin{equation}
M_{BH} ~\simeq~ 1.45 \times 10^5 M_{\sun} 
~\left ( {c\tau_{BLR} \over 1~{\rm light~day}} \right ) 
\left ( {\Delta v_{FWHM} \over 1000 {\rm ~km~s^{-1}}} \right )^2~.
\end{equation}
The size of the BLR can be estimated from (Kaspi et al. 1997; 
Wandel et al. 1999):
\begin{equation}
r_{BLR} ~=~ 15 L_{44}^{1/2} ~{\rm light~days}.
\end{equation}
where $L_{44}$ is the bolometric liminosity 
For MKN 421, estimates of the bolometric luminosity over the
$0.1$ to $1~\mu$m band vary considerably, and it is unclear whether
the quiescent or high-state luminosity is more appropriate. Using 
a bolometric luminosity derived from our HST magnitudes and the
conversion given in Laor (1998), we find $L_{44}\sim 40$, which
gives $r_{BLR} = 90$ light days. Assuming that $H\alpha$
is dominant in forming the broad line emission, then we use 
$\Delta v_{FWHM} \simeq 3300$ km s$^{-1}$ from our spectrum above. 
The implied black hole mass
is then $M_{BH}  ~\sim~  1.4 \times 10^{8}~ M_{\sun}$ which is of
of the same order of magnitude as the lower limit
estimate from the bulge mass above. 

A black hole mass of $10^9~ M_{\sun}$
is of order 0.25\% of the total dynamical mass out to 10 Kpc. Thus it is
not in itself enough to affect the large scale dynamics between the
two galaxies, although it would certainly strongly influence the inner bulge
regions. It is notable, however, that the center of mass of the system
will likely be significantly displaced from the position of the 
black hole, perhaps by a few hundred pc or more, depending on the
companion mass and the dark matter distribution. Careful measurements of
the virial motion of the stars near the center of MKN 421 may be able to
detect such an offset, which would provide 
independent confirmation of the proximity of MKN 421-5 to MKN 421.

\subsubsection{The dynamical state of the system}

The smooth rise in the rotation curve of MKN 421 out
to $R_p$ is consistent
with expectations for a post-encounter system, in which any major velocity
distortions produced by the encounter have relaxed and bulk tidal
motions now dominate. However we can not rule out the possibility that
this is the companion's first approach toward MKN 421. It is also
possible that there
are distortions in the velocity curves that are unresolved at our
present sensitivity.
Since we cannot unambiguously infer the companion's true velocity
from radial velocity measurements alone, we cannot prove that the
system is bound, and we can only base our estimates of the
tidal effects on the global properties we observe.
Despite these caveats, there are a number of different
lines of evidence that point to a bound and tidally relaxed system.

An obvious, though not compelling, piece of evidence is 
the fact that we observe the companion galaxy this close
to MKN 421 at all, at the relatively late epoch of the system. 
The probability that we are observing a first encounter between two high-ranking cluster galaxies cannot be estimated based on this single case alone.
However, based on the galaxy count of Ulrich (1978) and on our own Palomar
images over a 12 arcmin field around MKN 421, there are potentially
7 to 10 galaxies of comparable brightness to MKN 421-5, in a volume
of order $6 \times 10^{16}$pc$^3$, of order $10^4$ times the volume
containing MKN 421 and its companion. Since MKN 421 will probably tend to
significantly perturb the local potential, the chance of near encounters
with it are not determined simply by chance; however, we estimate
that the chance observation of such an encounter is of order 1\% or less
for any snapshot of similar clusters.

If the system is not in the initial phase of a first encounter,
it follows that (a) it is likely to be bound (due to the 
mass dominance of MKN 421 and the narrow range of phase space required
for MKN 421-5 to have the required escape velocity); and (b) that it
is tidally relaxed, since
the relaxation time for the system as observed is in the range of 100 Myr,
relatively short compared to the time of cluster formation.

The evidence for 
recent star-formation in the companion galaxy is also relevant to
discussion of the dynamics of this pair.
Recent simulations of hierarchical galaxy encounters 
(cf. Bekki 1999 \& references therein) have shown that 
star forming activity in the 
satellite galaxy is a likely result of such an interaction. A number of
recent HST studies have found a significant number of companions near 
QSO host galaxies (Bahcall et al. 1995; Disney et al. 1995) which appear to
be undergoing some type of interaction with the QSO host. Canalizo and Stockton
(1997) found that at least one companion galaxy in such a system
(PG 1700+518) showed evidence of a relatively young stellar
population.

What is perhaps surprising in the case of MKN 421 and MKN 421-5 is
that, although simulations such as those of Bekki (1999) indicate that
the companion galaxy is likely to evolve to a dwarf elliptical or
irregular as a result of the encounter, 
MKN 421-5 appears to in fact be a relatively normal spiral,
without any apparent major disruption or irregularity. This system thus
presents a challenge to models which cannot account for preservation of
such structure even in what was probably a strong tidal encounter
between these two systems. It is notable that MKN 421-5, as a probable Seyfert,
is likely to have a massive compact object in its nucleus. 
Based on arguments similar to the previous section and our
estimates of the bulge luminosity of MKN 421-5, the
expected mass is in the range of $2-5 \times 10^6~M_{\sun}$.
The presence of this large central mass may in fact have important dynamical
consequences in stabilizing a satellite galaxy in such an encounter.

\subsubsection{A schematic geometry for the system}

To postulate a possible three-dimensional geometry for this pair of 
galaxies, we 
combine a number of related pieces of evidence that bear on the
geometry, including our measured velocity curves, the apparent
inclination of the companion spiral, the major and minor
axes of MKN 421, and parameters associated with the central mass
of MKN 421, although the latter are included mainly for comparison
rather than as constraints on the global geometry.

The position angle of
the center of the companion relative to MKN 421 is $53^{\circ}$, 
and we estimate that the
projected major axis of the spiral lies at a position angle of $65^{\circ}$.
Thus the projected rotation axis (perpendicular to the major axis of the
spiral) of MKN 421-5
is at a position angle of $335^{\circ}$. The inclination of the spiral is
more difficult to estimate, since the shape may be confused by an
inner bar feature not easily distinguished in the images. However,
it appears that the spiral inclination is in the range of $45^{\circ}$
to $60^{\circ}$ to the line-of-sight. 

\begin{figure}
\plotone{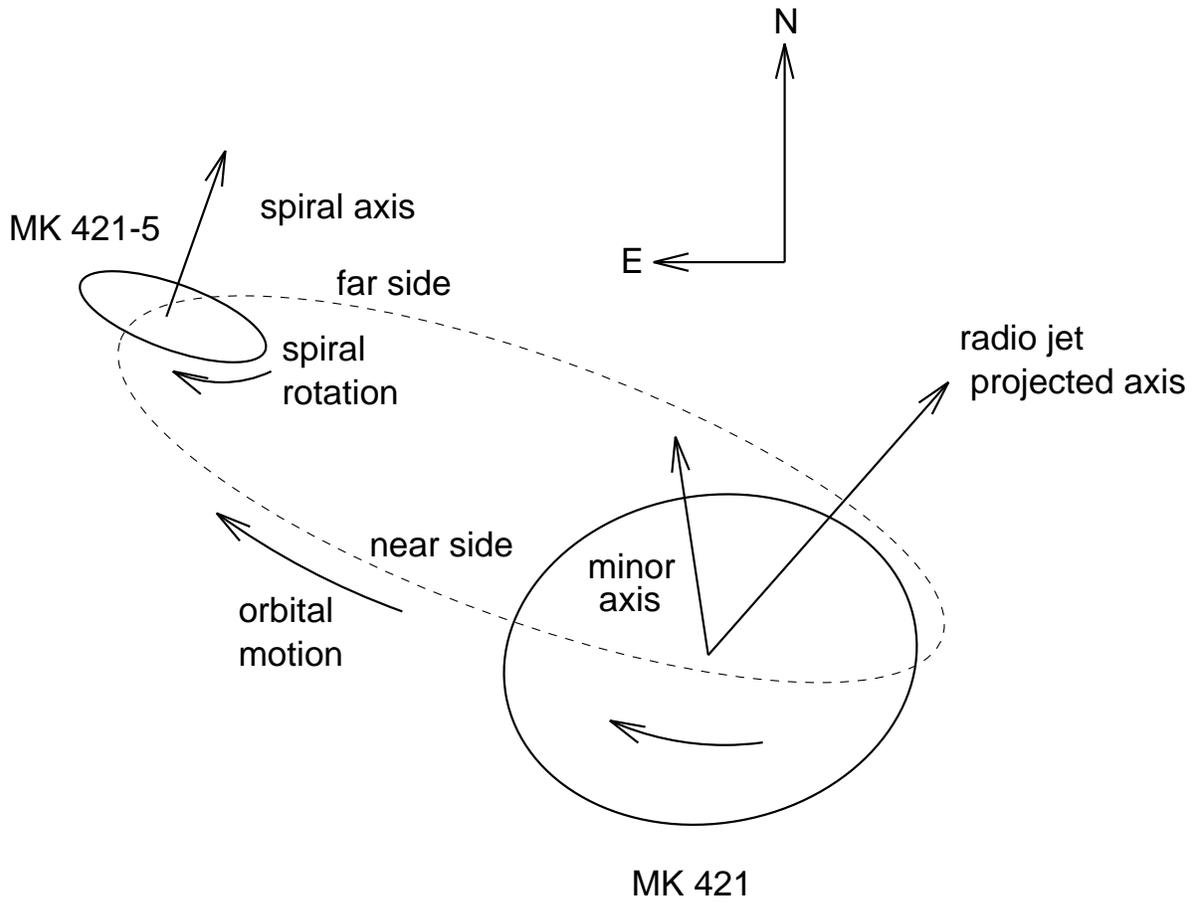}
\figcaption[Gorham.fig8.eps]
{A suggested schematic geometry for the 
MKN 421 + MKN 421-5 system. \label{mk421geom-fig} }
\end{figure}

If we assume that the disk of the spiral and the orbital plane are
roughly coincident (this is reasonable if the systems are
tidally relaxed), then we can make a first order schematic 
estimate of the overall geometry of the system, if the two galaxies are
bound in an orbit of only moderate eccentricity.
This schematic view of the system is depicted graphically in 
Figure \ref{mk421geom-fig}.
We have not attempted to quantify this geometry 
further; we present it simply as a qualitative baseline which can
be refined (or revised) by future observations.

Recent VLBI imagery and estimates of the
jet doppler factors from both VLBI and gamma-ray observations
(Piner et al. 1999; Gaidos et al. 1996) indicate conclusively 
that the jet angle must be $\leq 5^{\circ}$ with respect to the 
observation vector.
The projected position angle of the inner
jet from VLBI observations 
is $\sim 320^{\circ}$. We have plotted this position angle
in Fig. \ref{mk421geom-fig} for comparison. 
For the case of a small angle with respect
to the line-of-sight, the approximate coincidence of the
projected position angle with that of the spiral axis and the 
possible orbital plane is presumably an accident. However, if the jet
{\em is} actually physically aligned with the other angular momentum
vectors in the system, then a much larger angle for it with respect to the
line-of-sight is favored.

An outstanding question in MKN 421 and in BL Lac objects in general is
how well-aligned the jet axis is with the line-of-sight to the
AGN. If the results based on estimates of the doppler factors
are correct, and the jet axis is 
inclined by $\leq 5^{\circ}$, then we are led to conclude that either the
orbital plane of the two galaxies are not well-aligned with
the perpendicular to the jet axis, or
plane of the spiral companion is not well-aligned with the overall
orbital plane. In either case the residual torques may have
observable consequences, either in precession effects or tidal
velocity distortions of the system. More complete velocity mapping may
provide information on the latter consequence. As to the former
effect, further analysis is necessary to estimate whether there
might be observable effects, since the precession time scale would be
expected to be long compared to the orbital time scale.

\subsubsection{Companion rotation: Prograde or Retrograde?}

We conclude this section by briefly noting the sense of
the companion rotation.
The rotation curve shown in Fig. \ref{gradvel-fig} shows that
the eastern end of MKN 421-5 is redshifted relative to
the western end which is closest to MKN 421. This indicates a
rotation which is clockwise on the sky, and corresponds to
a prograde rotation relative to the projected rotation
of the bulge of MKN 421.

This feature of the system may
be important to the stability of the companion. Future simulations should
look in more detail 
at the effects of retrograde vs. prograde rotation in a tidal
encounter.

\section{Conclusions}

We have gathered evidence that supports the following conclusions:
\begin{enumerate}

\item MKN 421's nearest companion galaxy MKN 421-5 
appears to be an early-type spiral, rather than elliptical. 
Spectral evidence suggests moderately
recent star-formation activity in the companion, a feature which is
known to accompany galaxy interactions.

\item MKN 421-5 contains a Seyfert-like nucleus,
but without detectable emission lines in ground--based spectroscopy.
We find that its luminosity is probably too high for a
compact nuclear star cluster and conclude it is most likely to be an AGN,
in spite of the lack of spectroscopic evidence. 

\item We confirm the published radial velocity for MKN 421-5. We find that
MKN 421-5 is very likely to be bound to MKN 421, and its orbital
velocity is consistent with the trend in the bulk rotational velocity
of MKN 421, suggesting tidal interaction.

\item We report the first spectroscopic observation of emission lines
from MKN 421's nucleus, 
notably H$\alpha$ and NII. 
Both broad and narrow components of H$\alpha$ are present, with
the broad line emission showing a velocity dispersion of several
thousand km s$^{-1}$, typical of QSO emission lines.
Spectroscopic monitoring of these lines should continue.

\item The observed orbital velocity of the companion MKN 421-5
provides a lower limit on the mass
of MKN 421's bulge of $5.9 \times 10^{11}$ solar masses, with an estimated
bulge mass--to--light ratio of $\geq 17$. 

\end{enumerate}

\acknowledgements
We thank R. Linfield and the staff of Palomar Observatory for their 
invaluable help with the observations,
and Glenn Piner for his helpful comments on the manuscript.
This work was performed 
at the Jet Propulsion Laboratory, California Institute of Technology, under 
contract with the National Aeronautics and Space Administration. 
The National Radio 
Astronomy Observatory is a facility of the National Science Foundation, 
operated under a cooperative agreement by Associated Universities Inc.
This research is based in part 
on observations made with the NASA/ESA Hubble Space Telescope, 
obtained from the data archive at the Space
Telescope Science Institute. STScI is operated by the Association of Universities for Research in Astronomy, Inc. under
NASA contract NAS 5-26555.

\end{document}